%
%
%

\ifx\mnmacrosloaded\undefined \input mn\fi 
%
%
%
\input psfig
\newif\ifAMStwofonts

\ifCUPmtplainloaded \else
  \NewTextAlphabet{textbfit} {cmbxti10} {}
  \NewTextAlphabet{textbfss} {cmssbx10} {}
  \NewMathAlphabet{mathbfit} {cmbxti10} {} 
  \NewMathAlphabet{mathbfss} {cmssbx10} {} 
  \ifAMStwofonts
    \NewSymbolFont{upmath} {eurm10}
    \NewSymbolFont{AMSa} {msam10}
    \NewMathSymbol{\upi}     {0}{upmath}{19}
    \NewMathSymbol{\umu}     {0}{upmath}{16}
    \NewMathSymbol{\upartial}{0}{upmath}{40}
    \NewMathSymbol{\leqslant}{3}{AMSa}{36}
    \NewMathSymbol{\geqslant}{3}{AMSa}{3E}

     \let\le=\leqslant
     \let\ge=\geqslant
  \else
    \def\umu{\mu}
    \def\upi{\pi}
    \def\upartial{\partial}
  \fi
\fi



\loadboldmathnames



\letters          

%
%
%
%
\def\PBvp #1 #2{ #1, #2}
\def\PBa #1:#2 #3 #4 {#1,#2, {A\&A} \PBvp #3 #4}
\def\PBapj #1:#2 #3 #4 {#1,#2, {ApJ} \PBvp #3 #4}
\def\PBasupl #1:#2 #3 #4 {#1,#2, {A\&AS} \PBvp #3 #4}
\def\PBapjsupl #1:#2 #3 #4 {#1,#2, {ApJS} \PBvp #3 #4}
\def\PBpasp #1:#2 #3 #4 {#1,#2, { PASP} \PBvp #3 #4}
\def\PBpaspc #1:#2 #3 #4 {#1,#2, { PASPC } \PBvp #3 #4}
\def\PBmn #1:#2 #3 #4 {#1,#2, {MNRAS} \PBvp #3 #4}
\def\PBmsait #1:#2 #3 #4 {#1,#2, {Mem. Soc. Astron. It.} \PBvp #3 #4}
\def\PBnat #1:#2 #3 #4 {#1,#2, {Nat} \PBvp #3 #4}
\def\PBaj #1:#2 #3 #4 {#1,#2, {AJ} \PBvp #3 #4}
\def\PBjaa #1:#2 #3 #4 {#1,#2, {JA\& A} \PBvp #3 #4}
\def\PBaspsc #1:#2 #3 #4 {#1,#2, {Ap\&SS} \PBvp #3 #4}
\def\PBanrev #1:#2 #3 #4 {#1,#2, {ARA\&A} \PBvp #3 #4}
\def\PBrevmex #1:#2 #3 #4 {#1,#2, {Rev. Mex. de Astron. y Astrof.} \PBvp #3 #4}
\def\PBscie #1:#2 #3 #4 {#1,#2, {Sci} \PBvp #3 #4}
\def\PBesomsg #1:#2 #3 #4 {#1,#2, {The Messenger} \PBvp #3 #4}
\def\PBrmp #1:#2 #3 #4 {#1,#2, {Rev. Mod. Phys.} \PBvp #3 #4}
\def\PBans #1:#2 #3 #4 {#1,#2, {Ann. Rev. of Nucl. Sci.} \PBvp #3 #4}
\def\PBphrev #1:#2 #3 #4 {#1,#2, {Phys. Rev.} \PBvp #3 #4}
\def\PBphreva #1:#2 #3 #4 {#1,#2, {Phys. Rev. A} \PBvp #3 #4}
\def\PBphs #1:#2 #3 #4 {#1,#2, {Physica Scripta} \PBvp #3 #4}
\def\PBjqsrt #1:#2 #3 #4 {#1,#2, {J. Quant. Spectrosc. Radiat.
       Transfer} \PBvp #3 #4}
\def\PBcjp #1:#2 #3 #4 {#1,#2, {Can. J. Phys. } \PBvp #3 #4}
\def\PBjphb #1:#2 #3 #4 {#1,#2, {J. Phys. B} \PBvp #3 #4}
\def\PBapop #1:#2 #3 #4 {#1,#2, {Appl. Opt.} \PBvp #3 #4}
\def\PBgca #1:#2 #3 #4 {#1,#2, {Geochim. Cosmochim. Acta}\PBvp #3 #4}
%
%

%
%

%
%
%
%
%
\newcount\PBtn
\def\PBcleartn{\global\PBtn=0}
\def\PBtbl #1{\global\advance\PBtn by 1
\begintable{\the\PBtn} 
\caption{{\bf Table \the\PBtn .} #1}
}
\def\PBtbltwo #1{\global\advance\PBtn by 1
\begintable*{\the\PBtn} 
\caption{{\bf Table \the\PBtn .} #1}
}

\begintopmatter  

\title{Chemical abundances in the  young galaxy at z=2.309 towards PHL 957 }

\author{
  P. Molaro$^1$, M. Centuri\'on$^2$, G. Vladilo$^1$}
\affiliation{$^1$ Osservatorio Astronomico di Trieste, Via G.B. Tiepolo 11
34131, Trieste -- Italy}
\affiliation{$^2$ Instituto de Astrofisica de Canarias, Via Via Lactea, 
La Laguna, Tenerife Spain}
\shorttitle{The young galaxy towards PHL 957 }


\PBcleartn

\abstract { We present high-resolution  UES  spectra of the quasar PHL 957
obtained for studying the foreground Damped Ly$\alpha$  galaxy at z=2.309. 
Measurements of absorption lines lead to accurate
abundance determinations of  Fe, S and  N which complement  measurements of   Zn, 
Cr 
and Ni already available
for this system. We find [Fe/H]=$-2.0 \pm0.1$,  
[S/H]=$-1.54 \pm0.06$ and [N/H]=$-2.76 \pm0.07$.
 The ratio
 [Fe/Zn]=$-0.44$  provides evidence   that  $\approx$ 74\%  of iron and 
$\approx$ 28\% of zinc
are locked  into dust grains with a dust-to-gas ratio
of  $\approx$  3\% of the Galactic one. 
The total iron content in both gas and dust in the DLA system  is [Fe/H]=$-1.4$.
This   confirms 
 a rather low metallicity  in the galaxy, which    is in the early stages of its 
chemical evolution. The detection of SII allows us to measure 
the 
SII/ZnII ratio, which   is a  unique 
diagnostic tool for tracing back  its chemical history, since it is not affected 
by 
the presence of
dust. Surprisingly, the resulting relative abundance 
is  [S/Zn]=$0.0\pm0.1$,
at variance with  the  overabundance found in the  Galactic
halo stars with similar metallicity. We emphasize that the [S/Zn] ratio is solar 
in 
all the three DLA absorbers with extant data.   Upper limits are also found for  
Mn, 
Mg, O  and P and,  
once  the dust   depletion   is accounted for, we obtain   
[Mg/Fe]$_{c}$  $<$+0.2, [O/Fe]$_{c}$$<$0.4, [Mn/Fe]$_{c}$$<$+0.0 and 
[P/Fe]$_{c}$$<$$-$0.7. 
The [$\alpha$/Fe] values do not 
support a   Galactic  halo-like abundances  
implying  that the chemical evolution of this young galaxy is not
reproducing  our own Galaxy's evolution. 
} 


\maketitle  

\section{Introduction}
   
Abundances  in QSO absorption systems offer a unique opportunity to probe 
both 
chemical and dust evolution of galaxies at high redshifts.
Among the variety of the QSO absorbers 
the most useful for abundance determinations are the Damped Ly$\alpha$ (DLA) 
systems, since they provide very accurate   absolute measurements for the ion 
species  dominant in the  HI gas. The interest in the damped Ly$\alpha$ galaxies 
is 
amplified by the suggestion that they  are the likely 
 progenitors of the present day spiral galaxies  (Wolfe et al. 1986). 
 Research on chemical  abundances in  DLA systems was pioneered
by 
Black, Chaffee  \& Foltz   (1987) and 
Meyer \& York (1987).
The subject  has been recently reviewed by Lauroesch et al. (1996), to which
we refer for  detailed references. Recently a wealth of new data has been 
presented by Lu et al.  (1996) and Pettini et al.  (1997). The metallicities
 in the DLA absorbers  are found between $-$2.5 $\le$ [Fe/H] $\le  -0.5$,
and  largely overlap  those   found in the
 Galactic halo  or in  globular cluster stars 
 ([Fe/H]$ \equiv \log({Fe \over H}) - \log({Fe \over H})_{\sun}$).
 
The very early stages of the chemical
evolution of protogalaxies are   expected to be dominated by  
Type II supernovae  products  because   their   lifetimes are much shorter than 
those of 
 SNIa. 
 Type II SNe  nucleosynthesis is  characterized by  an enhancement
of $\alpha$ elements over the iron-peak elements, as   observed in the halo stars 
of 
the Galaxy, while the cumulative 
effects of both types of supernovae   essentially yield   solar ratios. In the 
Galaxy the transition  occurs 
at a metallicity of  [Fe/H]$\approx$ -1.0 corresponding  to a  few Gyrs after 
Galaxy
formation. If Damped Ly$\alpha$
are following 
a   similar chemical evolution  we should find a  halo-like  
pattern in high redshift systems (or [Fe/H]$<$ -1.0) and 
  a smooth   transition towards  solar composition at low redshift (or [Fe/H] 
$\ge$ 
-1.0).
 Several claims state  that the relative abundances 
 of DLA  systems are 
 consistent with the  halo pattern (Wolfe et al.  1994, Lu et al.  1996), while a 
change in the relative abundances at  low redshift may have been 
observed  for the first time in the candidate DLA  system at $z_{abs}$=0.558 
towards 
PKS 0118-272
  (Vladilo et al.  1997).
However,  DLA  abundances are measured in diffuse   interstellar  gas  and 
the intrinsic  abundances are  underestimated  if  some of the atoms are 
removed away from the gas phase and locked in dust grains. 
For instance, Si is generally found overabundant with respect to Fe 
 in  DLA absorbers, with
[Si/Fe] $\approx$ 0.4, but Si and Fe
 are differentially depleted from gas to dust
in our Galaxy and are  consistent   with the  observed  DLA overabundance.
 Probably
 dust   is  present as shown by  
 the reddening of QSO with damped systems (Pei, Fall \& Betchold, 1991) 
and by the observed [Zn/Cr] overabundance in  DLA systems
(Pettini et al.  1997). 
Therefore  the real chemical  pattern in the high redshift absorbers is not
yet firmly established.

In this paper we present new observations of the z=2.309 damped system towards
PHL 957 (V=16.6) with z$_{em}$ = 2.681. The system has been studied by 
 Meyer \& Roth (1990),
 Pettini, Boksenberg \& Hunstead (1990),
 Wolfe et al. (1994), and abundances have been derived for Cr, Zn, Ni.
Here we provide measurements for additional species such as S, Fe, N and  limits 
for 
O, Mn, P and Mg. It is shown that the relative ratios do not conform to a simple 
scaling of a Galactic  halo pattern in 
whatever way  modified by 
 dust.

\section{Observations  }

The data we present   are based on  observations
obtained with the Utrecht Echelle Spectrograph 
(Walker \& Diego 1985) at the Nasmyth focus 
of the 4.2 William Herschel Telescope at the 
Observatorio del Roque de los Muchachos on  La Palma island. 
Two sets of three spectra of 1 hour each were obtained in September 
and in December 1996, using a Tektronix CCD with 1024x1024, 
24$\mu$m square pixels for a total integration of $\approx$ 22000 s. 
The seeing was of $\approx$ 2 arcsec in both observing runs.
 The  slit 
width was set to 2.2 arcsec giving  a 4 pixel
projected slit  onto the detector that was binned by two pixels along the 
dispersion. 
The  resolving power   measured from  the
emission lines of the thorium-argon lamp frames 
is 
R = $\lambda$/$\Delta \lambda = 27000$ or $\Delta v \sim 11.1$
km s$^{-1}$ at all wavelengths.
We used the 31.6 grooves/mm grating 
which provides a  wavelength coverage of 
 $\lambda\lambda$ 3657-4617 \AA\ allowing
the  search for  the resonance 
transitions of  FeII $\lambda\lambda$ 1125.448 
  1127.098, 1133.665,   1143.226,    1144.938, 1260.533 \AA\,   
PII  $\lambda$   1152.818  \AA\, 
MnII   $\lambda\lambda$ 1197.184, 1199.391  \AA\        
MgII   $\lambda\lambda$ 1239.925, 1240.395   \AA\            
SII   $\lambda\lambda$  1250.584, 1253.811, 1259.519   \AA\   
OI    $\lambda\lambda$1302.168, 1355.598  \AA\
 NI    $\lambda\lambda$ 1134.980, 1134.415, 1134.165  \AA\
 at the redshift of $z=2.309$.
  The two
sets of spectra were reduced separately and then combined together 
using weights according to their S/N. 
  Cosmic ray removal, sky subtraction, 
optimal order extraction and wavelength calibration 
were performed  using the ECHELLE context  in MIDAS.
The rms scatter in the wavelength calibration was 0.7 km s$^{-1}$.
The two sets of data were normalized using a spline to connect smoothly 
the regions free from Ly $\alpha$ clouds. 
The final S/N is about 15 in correspondence of the SII lines and between 8 
and 11 for all the other lines. 

\beginfigure{1}
\psfig{figure=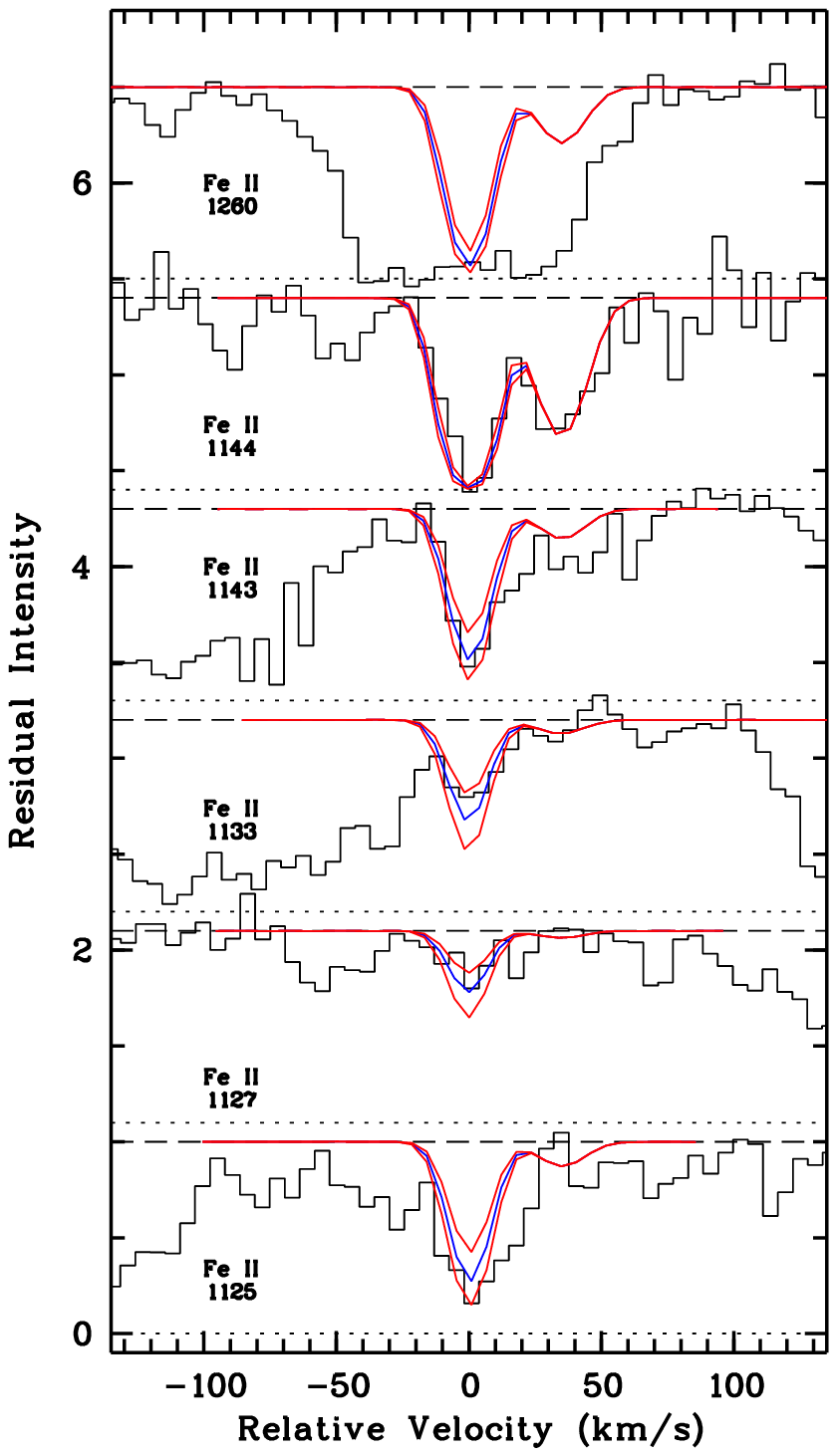,width=7.5cm}
\caption{Figure 1. Iron lines at z=2.309. 
Hystograms: observed spectra. Smooth lines: synthetic spectra.
The central smooth line results from the best fit  to
the FeII profiles  of the lower five panels. 
The FeII 1260 \AA\ line is  contaminated by 
Ly$\alpha$ lines. 
The upper and lower smooth lines encompass the $2\sigma$ 
column density error band.  
 }
 \endfigure

\subsection{ Column densities}

In Figs 1 and 2  we show the identified transitions  of FeII and SII
together with the spectral ranges of the resonance transitions of
MnII, MgII and PII.
In the figure the  zero of the  Doppler velocity scale
corresponds to $z = 2.30907$. A component at $v \sim 35 $ 
km s$^{-1}$ or $z = 2.30946$ is also  detected in  the stronger
transitions of  SII 1254~\AA\  and FeII 1144~\AA .  
Wolfe et al. (1994)  with  $\Delta v \sim $ 8 km s$^{-1}$  resolution
 spectra of  PHL957  
revealed an asymmetric profile of the $v$ = 0 km s$^{-1}$ main component,
suggesting the presence of an additional component at $v \sim 8$ km s$^{-1}$.

Column densities and b values were derived by a $\chi^2$ minimization 
of Voigt profiles convolved with the instrumental point spread function by 
using the FITLYMAN routine in MIDAS (Fontana \& Ballester 1995).
 Atomic parameters were taken from Morton (1991)
with the  exception of
the MgII  $\lambda$1240 \AA\ doublet, for which  the oscillator strengths were 
taken 
from Sofia, Cardelli \& Savage   (1994). The results of the fits 
are reported in Table 1 and shown in Figs 1 and 2 as a continuous line.
The effect  
of the double structure of the main component  on  the column densities of FeII 
and 
SII is negligible. In fact,
 using a two cloud model the  
fit of  FeII and SII lines  gives 
 a total column density which is   
  within 0.01 dex of the one-cloud model for both  elements.

\beginfigure{2}
\psfig{figure=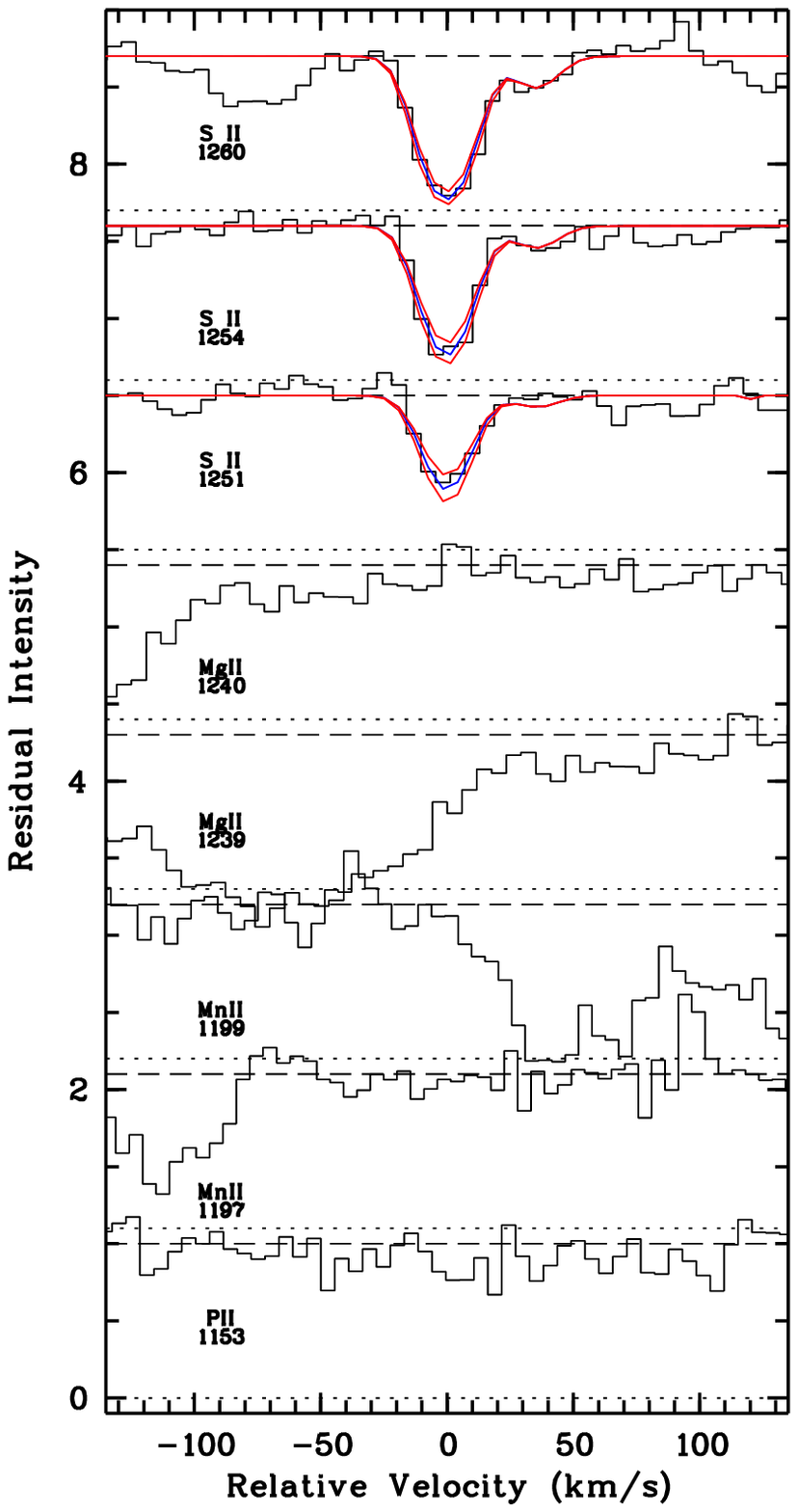,width=7.5cm}
\caption{Figure 2. Spectral regions of PHL 957 in correspondence of
the resonance lines specified in the labels. 
Hystograms: observed spectra. Smooth lines: synthetic spectra.
The central smooth line in the upper three panels
results from the best fit  to
the SII profiles.
The upper and lower smooth lines encompass the $2\sigma$ 
column density error band.  
 }
\endfigure

\PBtbl{Metallicities of the $z$=2.30907 damped Ly$\alpha$ system. 
}
\halign to \hsize{\tabskip=0ptplus12ptminus15pt  
# &
# &
# &
# &
# &
# &
#\cr
\multispan{7}{\hrulefill}\cr 
Ion  & $\log N(X) $ & \hfill $b$ \hfill& \hfill $[{X \over H}]$ &  \hfill [${X 
\over 
Fe}$] & \hfill $\delta_x$ \hfill &  \hfill [${X \over Fe}$]$_{c}$\cr
 \multispan{7}{\hrulefill}\cr
SII  & \hfill  15.07 $\pm$ 0.05 & \hfill 10.4  $\pm$ 0.5       & \hfill $-$1.54      
& 
       \hfill $+$0.45       & \hfill 0.00         & \hfill  $-$0.13           \cr
FeII & \hfill  14.89 $\pm$ 0.10 & \hfill 7.3  $\pm$ 0.7     &\hfill  $-$1.99      
&
       \hfill    ---        &    \hfill -0.58  &   \hfill ---         \cr
NI   & \hfill   14.69 $\pm$ 0.06   &  11.2 $\pm$ 1.0 \hfill             &\hfill  
$-$2.76      &
       \hfill  $-$0.77       & \hfill  0.00       &  \hfill $-$1.35            
\cr
PII  &   \hfill $< $12.76\hfill  &    & \hfill $<-$2.09  & 
       \hfill  $<-$0.10      &  \hfill 0.00       & \hfill $<-$0.68             
\cr
MnII &   \hfill $<$12.91  \hfill   &       & \hfill $<-$1.88  & 
       \hfill $<+$0.11       & \hfill -0.51    & \hfill  $<$$+$0.04         \cr
MgII &   \hfill $<$15.25 \hfill    &     & \hfill $<-$1.73  & 
       \hfill $<+$0.26      & \hfill -0.48       & \hfill  $<$+0.16        \cr
OI   &  \hfill $<$17.53  \hfill  &      & \hfill $<-$0.80  & 
       \hfill $<+$1.19      & \hfill 0.00         &  \hfill  $<$+0.61       \cr
\multispan{7}{\hrulefill}\cr
   }

\endtable

\section{ Discussion}
The elemental abundances or limits for the absorber at z=2.309 towards PHL 957
are reported in Table 1. The  adopted hydrogen column density is N(HI) = (2.5 
$\pm$ 
0.25) x 10$^{21}$  
 (Pettini, Boksenberg \& Hunstead 1990) and the solar abundances are from Anders 
\& 
Grevesse  (1989) except for Fe taken  from Hannaford et al. (1992).

The iron abundance  in the absorber  is [Fe/H]=$-$1.99, and when
combined  with the zinc value of [Zn/H]=$-$1.55 derived by Wolfe et al.  (1994), 
we 
obtain  [Fe/Zn]=$-$0.44.  In halo stars  zinc tracks  closely iron
 and  is almost undepleted  in the 
Galactic interstellar medium.  
The current interpretation for the Fe underabundance relative to Zn 
is that   the {\it missing} iron is  
tied up in dust grains.  An estimation of the elemental depletion in  DLA
systems for the various elements can be obtained by assuming
that the dust is not  different from that of our own Galaxy and with an
assumption on the intrinsic abundances. 
In the warm diffuse clouds of the Galaxy Fe and Zn are
depleted by -1.2 and -0.19  respectively 
(Savage \& Sembach 1996, Roth \& Blades 1996). Assuming that the intrinsic
ratio between Fe and Zn is solar   and 
scaling down the Galactic relative fraction of atoms 
of Fe and Zn 
 in dust grains to get
the observed [Fe/Zn]=-0.44, we infer  a depletion of -0.58 dex for Fe and -0.14 
dex 
for Zn in the DLA system towards PHL 957. This means that  $\approx$  74\%
of Fe and $\approx$ 28\% of Zn in PHL 957 is locked up in grains as opposed to 
the 
nearly total 
$\approx$ 94\% of Fe and the 35\%  of Zn in  the Galaxy.
For the other refractory elements Mg and Mn for which we present limits
 the predicted depletions
are -0.48 dex and -0.51 dex starting from a Galactic depletion of -0.82 dex 
and -0.92 dex respectively
(Savage \& Sembach 1996). Strictly speaking the depletions of Mg and Mn depend
on the  intrinsic 
abundances in   DLA absorbers. If Mg is overabundant the depletion factor
is smaller and if Mn is underabundant the depletion is  larger of what
 inferred here which holds for solar relative composition
 (Vladilo 1998). For the  elements S, N, P, and O we assume  null depletion 
as it is observed in the Galaxy.
The depletion factors and the
 abundances relative to Fe and corrected for dust contributions ($\equiv 
[X/Fe]_{c}$) are  reported in the last two columns of Table 1. 
These show that there is  a considerable reduced grain
condensation in the Damped system, although it is not negligible.
 If we consider  the grain contribution  the metallicity of the DLA 
system is 
[Fe/H]$_{c}$=-1.41, only slightly higher than that deduced  by the observed  
[Zn/H]=-1.55 ought to the small correction for depletion of Zn. This metallicity     
confirms previous indications  that the absorption system arises in a galaxy at 
an 
early stage
of its chemical evolution. 
With this value for the metallicity   the  dust-to-gas
ratio amounts to   about  3\%  of  that of the 
Galaxy.  This value is in agreement  with 
the values obtained by means of the [Cr/Zn] argument for this and other DLA 
absorbers   by  Pettini et al.  (1997).

Of particular importance 
is the comparison between $\alpha$ 
and iron peak elements.  
 In DLA systems 
the relative abundances of Si and Fe are often found
to be consistent with a halo-like pattern.
Lu et al. (1996) found an average value of 
$<{\rm [Si/Fe]}>$=+0.36$\pm$ 0.11 from a compilation of 12 measurements. 
However, Si and Fe are differentially depleted from gas to dust.
 The average value  of Galactic interstellar  clouds   is 
[Si/Fe]=+0.66$\pm0.26$ (Lu et al.  1995).
The observed enhancement of Si versus Fe in the DLA systems  would 
reflect the overabundances
of $\alpha$ elements with respect to the iron-peak elements only 
in  absence of dust.
Since DLA systems contain some amount of dust, as shown by the reddening of
QSO with DLA absorbers in their spectra (Pei, Fall \& Bechtold 1991), 
a moderate enhancement of Si over Fe cannot be regarded as  a
clear  evidence of a halo-like pattern.

 It has been pointed out  that   the ratio between
 SII and ZnII is probably the best 
 diagnostic tool available for revealing  the relative contributions from the 
different types of supernovae since S is mainly a product of type II SN,
while Zn of type Ia SN. This  because   both elements
show little affinity with dust  and 
their ratio is also safe against  possible contributions to the column densities
from HII regions along the line of sight, since they essentially cancel out
each other in the ratio (Molaro, Matteucci \& Vladilo  1995, Lauroesch et al.  
1996). Combining the Zn determination of Wolfe et al (1994) with our 
determination
for S we obtain [S/Zn]=0.01, or [S/Zn]$_c$=$-$0.13 when corrected for the small
Zn depletion, a result  which is at variance with the value 
[S/Fe]$\approx$ 0.5  observed in halo stars (Weeler, Sneden \& Truran 1989). In 
spite of the very low metallicity of the gas in this galaxy, 
 the  ratios of two elements  believed not to be significantly affected by
 dust depletion  are strictly solar. 
 Only few determinations for S and Zn
 are present in  literature.
Meyer \& Roth (1990) found [S/Zn]=$-$0.1
in the DLA system at $z=2.8$ towards 
QSO PKS 0528-250, and  this number is    confirmed by
 the new Keck measurements  by Lu et al.  (1996).
By combining  the column density published by   Kulkarni et al.  (1995) we obtain
 [S/Zn] = +0.1 
in the $z$=1.775 absorber in QSO 1331+170.
For these two cases a possible mild depletion of Zn in dust grains has not been
considered, and the ratios may even slightly decrease.
  Thus in all    systems with extant data for S and Zn  their ratio is found far 
from what observed in the Galactic halo stars.

The observed limit for the  $\alpha$ element Mg is [Mg/Zn]$\le$ $-$0.18 that
becomes  
 [Mg/Fe]$_{c}$$\le$ +0.16 when the Mg depletion
is applied. This value is not consistent with  
the typical overabundance of Mg observed in  the halo stars.

Oxygen abundance is not known  
since  the OI 1302.1685  \AA\ line is strongly  saturated and contaminated.  By 
using the non detection of the much fainter
 OI 1355.5977 line   we  
 obtain
[O/H]$<-$0.8 
which is consistent but somewhat less stringent than the  [O/H]$<-$0.97 derived 
by 
Wolfe et al.  (1994) at 3 $\sigma$ level. The limit by Wolfe
implies    [O/Fe]$_{c}$$<$ +0.44 and this is   only marginally consistent with
the O  enhancement of $0.5 <$ [O/H]$<$1   observed in the halo stars.

The upper limit for Mn yields a  ratio of
 [Mn/Fe]$_c$ $<$ +0.04.  The limit   is consistent with the moderate 
deficiency 
of $<{\rm [Mn/Fe]}>$=$-$0.32 $\pm$0.16
observed  by Lu et al.  (1996) 
in about 7 DLA absorbers.
The limit for the non-refractory 
phosphorus   is the first for a DLA absorber and gives [P/Fe]$_c$$<-$0.68.
 This   matches the expectations
for   an  odd-light element but the lack of P  observations
in metal poor stars   prevents tighter   considerations.

In our system we find [N/Fe]$_{c}$
 =$-$1.35.  
Nitrogen  is believed to be mostly a  
product of secondary nucleosynthesis,
but
 a primary component can be obtained when the
seed nuclei are produced in earlier 
helium burning stages of the same star. Nitrogen is essentially a non refractory 
element and dust problems are avoided but the  complex nucleosynthetic origin
 prevents a straightforward interpretation.  
 The nitrogen underabundance we find is significantly lower than
 that of halo dwarfs (Wheeler, Sneden \& Truran 1989), in line with the limit at 
[N/Fe]$<$-0.8 for the  absorber  at z=2.27936 towards 2348-147  (Pettini et al.  
1995),
but   at variance with the [N/Fe]$\approx$-0.39 
found 
in the   absorber   at z=3.390 towards QSO 0000-2619 (Molaro et al. 1996). 
 The measurements   show  that a real dispersion in the nitrogen abundances 
is probably  present among the DLAs, and we defer the discussion to a subsequent 
paper where we present other N observations in DLA systems (Centuri\'on et al.  
1997).

In summary, we find that all the $\alpha$ over iron ratios
[S/Fe]$_{c} \simeq -0.13$, 
[O/Fe]$_{c}$$<$+0.44 and [Mg/Fe]$_{c}$$<$ +0.16  are
  far  from the typical value of the halo. 
Assuming that the Zn and Fe track each other closely these results do not 
strongly rely on the depletion corrections we have applied, which have some 
intrinsic uncertainty related to the unknown
dust properties 
of these young galaxies.
The observed ratios  suggest that the star formation history
of this high redshift galaxy is different from that of the Milky Way.
Other  {\it anomalous} ratios have been observed in two other DLA  absorbers with 
solar S/Zn
  and in the DLA system at z=3.390 towards QSO 000-2619 where N 
has been  found at rather high levels (Molaro et al.  1996).
A project to extend S  observation to other DLA systems is currently
in progress. 

Alternative chemical evolution models for  explaining approximately
 solar-like ratios for $\alpha$ over iron-peak elements 
at
very low absolute abundances 
have been discussed  in Molaro et al.  (1995, 1996) and at length in Matteucci,
Molaro \& Vladilo (1997). These models adopt bursts 
 of star formation and/or selective galactic winds and are able to 
produce scenarios in which  Type I SNae and AGB   yields  are dominant.
 Such  models appear  
 suitable to interpret anomalous abundances such  those 
found in the z=2.309 DLA absorber towards PHL 957.  If  chemical patterns 
different 
from the Galactic halo should be 
  found either common or frequent  among the Damped  Ly$\alpha$ galaxies, through 
future observations  there would be important implications
for  disclosing the   nature of the galaxies which are responsible for the  DLA 
systems. The present observations show that at least few DLA absorbers
do not conform to the notion that they are the   progenitors of the
present day spiral galaxies.


\section*{References}

\beginrefs
\bibitem Anders E., and Grevesse N., 1989, Geochim. Cosmochim.
Acta 53, 197

\bibitem  
Black J.H., Chaffee F.H., Foltz C.B., 1987, ApJ, 317, 442
 \bibitem  
Centuri\'on M., Bonifacio P., Molaro P., Vladilo G., 1997 in prep.
 
\bibitem  
 Fontana A., Ballester P., 1995 The Messenger 80, 37
 

\bibitem Hannaford P., Lowe, R.M., Grevesse N., Noels, A., 1992 A\&A, 259, 301

 \bibitem  
Kulkarni V.P., Huang K., Green R.F., Bechtold J., Welty D.,
York D.J., 1996, MNRAS,  279, 197
 
\bibitem  
 Lauroesch J. T., Truran, J. W., Welty D. E., York D. G. 1996, PASP, 108, 641

\bibitem  
 Lu L., Savage B.D., Tripp T.M., Meyer D.M. 1995, ApJ 447, 597 

\bibitem  
 Lu L., Sargent L.W., Barlow T. A. Churchill C. W., Vogt S. S. 1996 ApJS, 107, 
475

\bibitem  
 Meyer D.M., Roth K.C., 1990, ApJ, 363, 57
 
 
\bibitem  
 Meyer D.M., York D.G., 1987, ApJ, 319, L45

  
\bibitem  
 Matteucci F., Molaro P., Vladilo G., 1997, A\&A, 321, 45

 \bibitem  
 Molaro P., Matteucci F., Vladilo G., 1995,   proc.  
"Observational Cosmology: from Galaxies to Galaxy Systems" Sesto 4-7 July.


\bibitem  
 Molaro P., D' Odorico S., Fontana A., Savaglio S., Vladilo G.,
1996, A\&A, 308, 1


 \bibitem  
Morton D.C.,  1991 ApJS, 77, 119


 \bibitem  
Pei Y.C., Fall S. M., Bechtold, J. 1991, ApJ, 378, 6 

\bibitem  
Pettini M., Boksenberg A., Hunstead R.W., 1990, ApJ, 348, 48

 
\bibitem  
 Pettini M., King D.L., Smith L.J., Hunstead R.W., 
1997, ApJ, 478, 536
 
 
 
\bibitem
Roth K. C., Blades J. C., ApJ, 445 L95

 \bibitem  
Savage B. D., and Sembach K. R., 1996, ARA\&A, 34, 279


 \bibitem  
Sofia U. J., Cardelli J. A.,  Savage B. D., 1994, ApJ 430, 650
 

\bibitem  
 Vladilo G, Centuri\'on M., Falomo R., Molaro P., 1997, A\&A 327, 47

\bibitem
 Vladilo G., 1998, ApJ, February 1st
 
 \bibitem  
Walker D. D.,  \& Diego F., 1985, MNRAS, 217, 355

\bibitem  
Wheeler J.C., Sneden C., Truran J.W., 1989, A\&AR, 27, 279

 \bibitem  
Wolfe A.M., Turnshek D.A., 
Smith H.E., Cohen R. D., 1986, ApJS, 61, 249 

\bibitem  
 Wolfe A. M., Fan X. M., Tytler D., Vogt S. S., Keane M. J., Lanzetta K. M.
1994, ApJ, 435, L101

\bibitem  
 Wolfe A. M. Lanzetta K. M. Foltz C. B. Chaffee F. H. 1995, ApJ, 454, 698

\endrefs
\bye